\documentclass[preprint,aps,floats,showpacs,amssymb,tightenlines]{revtex4}
\usepackage{epsfig}

\def\half{\textstyle{1\over2}}

\newcommand{\be}{\begin{equation}}
\newcommand{\ee}{\end{equation}}
\newcommand{\bea}{\begin{eqnarray}}
\newcommand{\eea}{\end{eqnarray}}
\newcommand{\bml}{\begin{mathletters}}
\newcommand{\eml}{\end{mathletters}}

\begin{document}

%%%%%%%%%%%%%%%%%%%%%%%%%%%%%%%%%%%%%%%%%%%%%%%%%%%%%%%%%%%%%%%%%%%%%%%%%%

\title{Exotic composites: the decay of deficit angles in global-local monopoles}
\renewcommand{\thefootnote}{\fnsymbol{footnote}}
\author{Ana Ach\'ucarro\footnote{achucar@lorentz.leidenuniv.nl}}
\affiliation{Lorentz Institute of Theoretical Physics, Leiden University, 2333 RA Leiden, The Netherlands}
\affiliation{Department of Theoretical Physics, University of the Basque Country UPV-EHU, 48080 Bilbao, Spain} 
\author{Betti Hartmann\footnote{b.hartmann@iu-bremen.de}}
\affiliation{School of Engineering and Science, International University Bremen (IUB),
28725 Bremen, Germany}
\author{Jon Urrestilla\footnote{kap10@sussex.ac.uk }}
\affiliation{Department of Physics and Astronomy, University of Sussex, Brighton, U.K.   }
\date{\today}

%%%%%%%%%%%%%%%%%%%%%%%%%%%%%%%%%%%%%%%%%%%%%%%%%%%%%%%%%%%%%%%%%%%%%%%%%%
\begin{abstract}
We study static, spherically symmetric, composite
global-local monopoles with a direct interaction term between the two
sectors in the regime where the interaction potential is large.  At
some critical coupling the global defect disappears and with it the
deficit angle of the space-time. We find new solutions which represent
local monopoles in an Anti-de-Sitter spacetime. In another parameter
range the magnetic monopole, or even both, disappear. 
The decay of the magnetic monopole is accompanied by a dynamical transition from the higgsed phase to
the gauge-symmetric phase. We comment on the applications to cosmology, 
topological inflation and braneworlds.
\end{abstract}

\pacs{04.20.Jb, 04.40.Nr, 11.27.+d, 98.80.Cq}
\maketitle

%%%%%%%%%%%%%%%%%%%%%%%%
\section{Introduction}
%%%%%%%%%%%%%%%%%%%%%%%%

The properties and interactions of branes and D-strings have few
analogues in field theory and those analogues are by no means perfect.
In this respect, the interplay of global and local symmetries can
provide very useful toy models \cite{DV02,DV03}.  Here
we consider one such example in which issues like the decay of deficit
angles, gauge symmetry restoration and a dynamical 
reduction of the effective cosmological constant (as would be needed,
for instance, to exit inflation) can be investigated analytically and
numerically. The model has $SU(2)_{global} \times SU(2)_{gauge}$
symmetry with a direct coupling between the two scalar sectors and in
a certain range of parameters it admits a composite defect made of a
magnetic monopole and a global monopole (a hedgehog).

Defects consisting of a local monopole and a global monopole
interacting solely through gravity were considered by Olasagasti
\cite{O} in the $\sigma$-model approximation. 
The full field theoretical solutions coupled to gravity were given in
\cite{spi,bbh} for the case of zero or weak coupling, respectively.
Here we investigate the limit of strong coupling.

The stability of such composite systems is unclear \cite{AU,B} but
here we take the point of view, inspired by braneworlds, that the
cores are fixed at one particular point in space (this would be the
case, for instance, if the branes were localized at the minimum of
some potential in the higher dimensional space). We investigate the
static, spherically symmetric solutions.

Magnetic monopoles were first introduced by 't Hooft and Polyakov
\cite{thooft}. In flat space they
have a size roughly of order the inverse vector mass $e\eta_1$, where
$e$ is the SU(2) coupling constant and $\eta_1$ is the v.e.v. of the
Higgs field. The monopoles carry a quantised  
magnetic charge   $2\pi n /e$, where $n$ is the winding number
of the monopole. The 't Hooft-Polyakov monopole
in curved space-time was first studied in \cite{br}, and a  
Reissner-Nordstr\"om type solution
to the coupled system of equations was found. After \cite{nwp} gave hints
that globally regular, gravitating monopole solutions should exist,
these solutions were constructed numerically in \cite{ortiz, lnw, bfm}.
It was shown that these gravitating monopole solutions
exist only for small monopole masses. When the Schwarzschild radius of
the solution $\propto \eta_1 G/e$ (where $G$ is Newton's constant)
is comparable to the size of the monopole $\propto (\eta_1 e)^{-1}$, the monopoles
become black holes which outside of the horizon correspond to the
Reissner-Nordstr\"om solution and are thus uniquely characterised
by their charge and mass. On the other hand, mini-black holes
sitting inside the core of the monopoles are also possible \cite{lnw,bfm}.
These ---in contrast to the Reissner-Nordstr\"om solutions---
 have non-trivial matter fields outside their horizon and thus violate
the no-hair conjecture.

In flat space, global monopoles have a logarithmically divergent
energy coming from the gradients of the scalar field far from the
core, and their stability with respect to angular collapse has been a
matter of confusion for some years \cite{globalstable}, possibly
because global monopoles are stable to infinitesimal axisymmetric
perturbations, but the spherical configuration can be deformed with
finite {\it extra} energy to a new decay channel \cite{AU00}. In
\cite{wt}, it has been demonstrated that the gravitating global
monopoles are stable against spherical as well as polar perturbations.
When gravitational effects are considered, as shown by Barriola and
Vilenkin, global monopoles have a solid deficit angle \cite{vilenkin2}
$\delta \propto G \eta_2^2 $ if $\eta_2$ is the vacuum expectation
value of the scalar field forming the monopole. In this
paper we consider solutions with deficit angle smaller than
$4\pi^2$ ($8 \pi G \eta_2^2 < 1$). For larger deficit angles there are no static solutions, and
in particular there can be {\it topological inflation} at the monopole
core \cite{Vilenkin,Linde}.

The composite  local-global monopole solutions with spherical symmetry were
studied in detail by Spinelly et al \cite{spi}. 
The spherically
symmetric composite monopoles show a number of interesting features:
in \cite{spi} the system with only ``indirect'' interaction
via gravity was studied. It was found that far from the core of the composite
defect, the space-time corresponds to a Reissner-Nordstr\"om space-time with
solid deficit angle. 

An interaction term between the Higgs field of the magnetic monopole
and the scalar field that forms the global monopole was introduced
shortly afterwards, and again the spherically symmetric solutions were
%analysed in \cite{bbh,bh} in the case of weak interaction. The purpose
of this paper is to show that, when the interaction term is
sufficiently large, there is a major qualitative change in the nature of
the defects and the surrounding spacetime as the defects become unstable  and
disappear.

The basic idea in the monopole model analysed below is the following.
The model contains gravitational and gauge fields and two sets of
scalars. The defect is a composite defect made of a topological global
monopole giving a solid deficit angle to the spacetime around it, and
a topological gauged ('t Hooft- Polyakov) one giving a long-range
magnetic field.  The vacuum manifold is $S^2 \times S^2$. An
interaction term between the two sets of scalars is introduced. 
The interaction term couples the scalars and is such that the
composite defect is in a spacetime with no cosmological
constant (other choices are possible and they will be
considered elsewhere).

For low coupling, the main
changes are to the detailed profiles and spacetime structures around
the defect, and were analysed in \cite{bbh}. In particular the mass
that one assigns to the composite becomes positive in certain regimes
(as opposed to negative for the bare global monopole).

For critical values of the scalar couplings, however, a drastic
change occurs as the vacuum manifold changes to $S^5$ and
the monopoles become non-topological. At this point it becomes
energetically favourable for the global monopole to decay, thus
removing the linearly divergent energy due to the slow fall-off of
scalar gradients far from the core. 
The solid deficit angle of the solution disappears 
with the global monopole, and the 't Hooft-Polyakov monopole
changes its core structure and mass (but keeps the same magnetic
charge).

The interaction coupling can be increased further. In that case we
find two possibilities, depending on parameters: either we
recover the AdS monopole solutions of \cite{lms,ls} --that is, the stable
defects are magnetic monopoles in a spacetime with negative
cosmological constant-- or the system may also be able to decay to the
true vacuum and thus get rid of both monopoles.

Our paper is organised as follows: in Section II, we give the model including
the spherically symmetric Ansatz, the equations of motion and the boundary
conditions. In Section III, we discuss analytic results for 
the defects arising in the limit of large coupling. In Section IV, we present
our numerical results and in Section V, we summarize and discuss our results.

%%%%%%%%%%%%%%%%%%%%%%%%%%%%%%%%%%%%%%%%%%%%%%%%%%%%%%%%%%%%
\section{The Model}
%%%%%%%%%%%%%%%%%%%%%%%%%%%%%%%%%%%%%%%%%%%%%%%%%%%%%%%%%%%%

The action of the composite defects' system that we will consider
reads \cite{bbh}:
\begin{equation}
\label{action}
S=S_{G}+S_{M}=\int d^4 x \sqrt{-g} \left( \frac{1}{16\pi G} R + {\cal L}_{local}
+{\cal L}_{global}+{\cal L}_{inter}\right)  \ ,
\end{equation}
where $R$ is the Ricci scalar, $G$ is Newton's constant
and ${\cal L}_{local}$, ${\cal L}_{global}$ and ${\cal L}_{inter}$
denote, respectively, the Lagrangian density of the 
local defect, the global defect and the interaction potential, which
couples the two sectors directly.

We have:
\begin{eqnarray}
{\cal L}_{local}&=&-\frac14 F_{\mu\nu}^a F^{\mu\nu,a}-\frac12(D_\mu\phi^a)
(D^\mu\phi^a)-\frac{\lambda_1}{4}\left(\phi^a \phi^a -\eta_1^2\right)^2 \ , \nonumber \\
{\cal L}_{global}&=&-\frac12(\partial_{\mu} \chi^a) (\partial^{\mu} \chi^a) -
\frac{\lambda_2}{4}\left(\chi^a \chi^a -\eta_2^2\right)^2
\end{eqnarray}
and
\begin{equation}
\label{interm}
{\cal L}_{inter}= -
\frac{\lambda_3}{2}\left(\phi^a \phi^a -\eta_1^2\right)\left(\chi^a\chi^a-\eta_2^2\right) \ ,
\end{equation}
with $a=1,2,3$, and $\phi^a$ and $\chi^a$ are scalar triplets.  This
makes the total potential the most general quartic potential 
invariant under $SU(2)_{gauge} \times SU(2)_{global}$ 
up to an additive constant whose effect is to change the value of the
cosmological constant. Here we choose the constant to be zero when
$|\phi^a| = \eta_1, \ |\chi^a| = \eta_2$, which means that in the range
of parameters where the composite monopole solution exists, the
cosmological constant is zero.

The field strength tensor and covariant derivative of the Higgs field read: 
\begin{equation}
F_{\mu\nu}^a=\partial_\mu A_\nu^a-\partial_\nu A_\nu^a-e\epsilon_{abc}
A_\mu^b A_\nu^c \ \ , \ \ \  
D_\mu\phi^a=\partial_\mu\phi^a-e\epsilon_{abc}A_\mu^b\phi^c \ ,
\end{equation}
where $A_\mu$ is a $SU(2)$ gauge field and  $e$ is the gauge coupling constant.

%%%%%%%%%%%%%%%%%%%%%%%%%%%%%%%%%%%%%%%%%
\subsection{The structure of the vacuum manifold}
%%%%%%%%%%%%%%%%%%%%%%%%%

Consider the scalar potential 

\be V(\phi,\chi) = \frac{\lambda_1}{4}\left(\phi^a \phi^a
-\eta_1^2\right)^2 +\frac{\lambda_2}{4}\left(\chi^a \chi^a
-\eta_2^2\right)^2 +\frac{\lambda_3}{2}\left(\phi^a \phi^a
-\eta_1^2\right) \left(\chi^a\chi^a-\eta_2^2\right) \ee

Global stability of $V$ requires $\lambda_1 >0, \
\lambda_2>0$. $\lambda_3$ can have either sign, but
stability requires $\lambda_3 > -\sqrt{\lambda_1 \lambda_2}$. 

When $\lambda_3>0$ and  $\Delta = \lambda_1\lambda_2 - \lambda_3^2 > 0$ 
the vacuum manifold is the same as in the non-interacting ($\lambda_3=0$) case,
$S^2\times S^2$, given by 
\be \phi^a\phi^a = \eta_1^2 \ , \qquad
\chi^a\chi^a = \eta_2^2 
\label{etas}
\ee 
This case was analysed in great detail in
\cite{bbh}.  

However, if $\Delta = \lambda_1\lambda_2 - \lambda_3^2 = 0$ $(\lambda_3>0)$ the vacuum
manifold becomes topologically equivalent to $S^5$ as can be seen from
the identity \be V = \frac{1}{4} \left[ \sqrt{\lambda_1} (\phi^a\phi^a -
\eta_1^2) + \sqrt{\lambda_2} (\chi^a\chi^a - \eta^2) \right]^2 +
\frac{1}{2} [ \lambda_3 -\sqrt{\lambda_1\lambda_2}]
\left(\phi^a\phi^a-\eta_1^2\right)\left(\chi^a\chi^a-\eta_2^2\right)  \ .
\ee
The second term is identically zero and the vacuum states are
now given by the condition \be \sqrt{\lambda_1} \phi^a\phi^a +
\sqrt{\lambda_2} \chi^a\chi^a = \sqrt{\lambda_1} \eta_1^2 +
\sqrt{\lambda_2} \eta_2^2  \ .  \ee
The monopoles are no longer topological. We call this the {\it critical coupling} case.

We can consider even larger values of $\lambda_3$. In that case,
$\lambda_3 >\sqrt{\lambda_1\lambda_2} $,  there are two possible minima
of the potential 
\be
\label{minimum1}
\phi^a = 0 \ , \quad  \chi^a\chi^a   = \eta_2^2+\eta_1^2 \frac{\lambda_3}{\lambda_2} \quad {\rm with} \quad V_f=\frac{\eta_1^4}{4\lambda_2} \Delta
\ee
and
\be
\label{minimum2}
\phi^a\phi^a = \eta_1^2 + \eta_2^2 \frac{\lambda_3}{\lambda_1} \ , 
\quad \chi^2 = 0 \quad {\rm with} \quad V_h=\frac{\eta_2^2}{4\lambda_1}\Delta
\ee

In this regime the composite monopole is unstable.  The potential has
two minima with different (negative) values, and we might expect the
lowest one to be favoured if there is no winding.  In general,
(\ref{minimum1}), respectively (\ref{minimum2}) is the lowest minimum
for $\eta_1^4 \lambda_1 > (<) \ \eta_2^4 \lambda_2$.  However, the
potential energy ``criterion'' is not enough to predict
what will be the static solution. Actually, numerical simulations show
that the static solution has $\chi^a=0$ for all cases, due to the
linearly divergent gradient energy of the global monopole (even at
$\Delta=0$). So the global monopole disappears for all choices of
parameters.  The solution then is a static solution corresponding to a
magnetic monopole in Anti-de-Sitter space (see below for more
details). The monopole core size and asymptotic field values change
according to the new vacuum expectation value of the Higgs field.

For the case $\lambda_3<0$ and $\Delta=\lambda_1\lambda_2-\lambda_3^2>0$ the situation is
analogous to the previous case. 
The solutions tend  asymptotically to
(\ref{etas}), with some changes in the profile, 
as we will show in section \ref{numerical}.

On the other hand for $\lambda_3<0$ and $\Delta=0$, the
potential becomes \be V = \frac{1}{4} \left[ \sqrt{\lambda_1}
(\phi^a\phi^a - \eta_1^2) - \sqrt{\lambda_2} ( \chi^a\chi^a - \eta_2^2) \right]^2
\label{fi}
\ee

with minima
\be \sqrt{\lambda_1} \phi^a\phi^a -
\sqrt{\lambda_2} \chi^a\chi^a = \sqrt{\lambda_1} \eta_1^2 -
\sqrt{\lambda_2} \eta_2^2  \ .
\label{l3n}  \ee

The existence of different solutions depends strongly on the values of 
the parameters. Contrary to the previous
case, it might be possible that the global monopole
``survives'', while the local monopole disappears. Let us be more specific:

If $\sqrt{\lambda_1}\eta_1^2<\sqrt{\lambda_2}\eta_2^2$, then the global monopole will not disappear, because
equation (\ref{l3n}) cannot be fulfilled for $\chi=0$; but the local one can disappear.  
Likewise, the possibility of having the global monopole disappear and the local remain persists for  
$\sqrt{\lambda_1}\eta_1^2>\sqrt{\lambda_2}\eta_2^2$.

This is somewhat analogous to what happens in
D-term N=1 supersymmetric models with a Fayet-Iliopoulos (FI) term. Introducing  $\kappa_{\rm
FI}\equiv\sqrt{\lambda_1}\eta_1^2-\sqrt{\lambda_2}\eta_2^2$, 
equation (\ref{fi}) can be rewritten suggestively as \be V=\frac14 \left[
\sqrt{\lambda_1}|\phi|^2 - \sqrt{\lambda_2} |\chi|^2 - \kappa_{\rm FI}
\right]^2 \\ .  
\ee 
Static defects (vortices) in models with such a
potential were studied in \cite{PRTT,ADPU}. One of
the fields is identically zero everywhere, and the field that
vanishes is given by the sign of $\kappa_{\rm FI}$: if $\kappa_{\rm
FI}>0$ then $\chi=0$; and if $\kappa_{\rm FI}<0$ then $\phi=0$.

When the magnetic monopole decays, the magnetic field becomes weaker
at the core and the core expands to an infinite size. At any stage of
the decay, the total magnetic flux integrated on a sphere sufficiently
far from the monopole core is constant, but there is no  localized 
object
carrying this magnetic flux. The end result is indistinguishable from
Minkowski space except for the conserved magnetic flux that has been
diluted to an unobservable level. (The situation is completely
analogous to the semilocal string model in the unstable regime,
where there is a conserved flux which nevertheless spreads over an
infinite area and is not carried by any localized  vortex \cite{VA91,H92,P92}).

There is a boundary case
$\sqrt{\lambda_1}\eta_1^2=\sqrt{\lambda_2}\eta_2^2$, where {\bf both}
monopoles disappear at the same time, and the magnetic charge gets
diluted (the analogous case would be supersymmetric QED without FI
term, where the possible vortices were shown to be unstable and decay
to the vacuum \cite{ARH}).

%%%%%%%%%%%%%%%%%%%%%%%%%%%%%%%%%%%%%%%%%
\subsection{Spherically symmetric Ansatz}
%%%%%%%%%%%%%%%%%%%%%%%%%%%%%%%%%%%%%%%%%%

In what follows we consider configurations with spherical
symmetry, with the two monopole cores fixed at the origin. It is not
obvious that this assumption is justified dynamically, since the
stability of the composite monopole is still unclear \cite{AU,B} and
it is possible that the cores of the two monopoles will expel each
other, separating the global and local monopoles. Nevertheless we are
interested in a possible application to braneworlds, where the higher
dimensional theory might provide a pinning potential (not manifest in
the low energy theory) which keeps the cores at the origin \cite{foot}.

The Ansatz for the spherically symmetric metric tensor in
Schwarzschild-like coordinates reads:
\begin{equation}
ds^2=-A^2(r)N(r)dt^2+N^{-1}(r)dr^2+r^2(d\theta^2+\sin^2\theta d\phi^2) 
\end{equation}
where we define for later convenience the mass function $m(r)$ as follows~:
\begin{equation}
\label{N}
N(r)=1 -8\pi G \eta_2^2 -\frac{2m(r)}{r} \ . \ \
\end{equation}
Note that in the case of spherical
symmetry the only choice for the winding number is $n=1$ or $n=0$.
Higher winding number configurations would have axial or even less
symmetry. We consider a composite monopole with unit winding in both the global
and local sectors. The Ansatz for the static spherically symmetric global field
$\chi^a$, the Higgs field $\phi^a$ and the gauge field $A_{\mu}^a$ in
Cartesian coordinates reads:
\begin{equation}
\phi^a(x)=\eta_1 h(r)\hat{x}^a \ , \ \  \chi^a(x)=\eta_1 f(r)\hat{x}^a \ ,
\end{equation}
\begin{equation}
A_i^a(x)=\epsilon_{iaj}\hat{x}^j\frac{1-u(r)}{er} \ , \ \ A_0^a(x)=0 \ .
\end{equation}

For convenience 
we have rescaled the global field 
with the vacuum expectation value of the local field. 
Note that the configuration given by a $n=1$ magnetic monopole and
\begin{equation}
\label{condensate}
\chi^a = \eta_1 f(r) {n_\chi}^a \qquad {n_\chi}^a {n_\chi}^a = 1 \ ,
\label{es1}
\end{equation}
with $n_\chi^a $ a fixed direction in internal space, is also spherically
symmetric. Similarly, the $n=1$ global monopole 
\begin{equation}
\phi^a(x)=\eta_1 h(r){n_\phi}^a \ , \ \ 
A_\mu^a(x)= 0 \, \ ,
\label{es2}
\end{equation}
with ${n_\phi}^a$ fixed, is also spherically symmetric. The same is true of the ansatz given by the previous two equations (\ref{es1},\ref{es2}), which has zero winding in both sectors.

%%%%%%%%%%%%%%%%%%%%%%%%%%%%%%%
\subsection{Equations of motion}
%%%%%%%%%%%%%%%%%%%%%%%%%%%%%%%%%
Varying (\ref{action}) with respect to the matter fields and gravitational
fields and introducing the dimensionless variable $x$ and dimensionless mass function
$\mu(x)$ \cite{bbh}:
\begin{equation}
x=e\eta_1 r \ , \ \mu(x)=e\eta_1 m(r) 
\end{equation} 
as well as the dimensionless coupling constants:
\begin{equation}
q=\eta_2/\eta_1 \ , \ \ \ \gamma^2=8\pi G\eta_1^2 \ , \ \ \  \beta_i=\lambda_i/e^2 \ , \
i=1, 2, 3
\end{equation}
we obtain the following set of differential equations (the prime denotes the
derivative with respect to $x$): 
\begin{equation}
\left[ x^2AN h'\right]'=A\left[2u^2h+x^2\beta_1 (h^2-1)h
+{x^2\beta_3 }h(f^2-q^2)\right] \ ,
\label{eqnh}
\end{equation}
\begin{equation}
\left[x^2AN f'\right]'=
A\left[2f+x^2\beta_2 f(f^2-q^2)+
{x^2\beta_3 } (h^2-1)f\right] \ ,
\label{eqnf}
\end{equation}
\begin{equation}
\left[AN u'\right]'=A\left[\frac{u(u^2-1)}{x^2}+
uh^2\right] \ ,
\label{eqnu}
\end{equation}
\begin{equation}
\label{xAN}
(xAN)'=[1-\gamma^2x^2\bar{U}]A \ ,
\end{equation}
\begin{equation}
A'=\gamma^2 Ax\bar{K}
\label{eqnA}
\end{equation}
where we have defined
\begin{eqnarray}
\bar{U}&=&\frac{(u^2-1)^2}{2x^4}+\frac{u^2h^2}{x^2}+\frac{f^2}{x^2}+
\frac{\beta_1}4(h^2-1)^2+\frac{\beta_2}4(f^2-q^2)^2\nonumber\\
&+&\frac{\beta_3}2(h^2-1)(f^2-q^2) \ ,
\end{eqnarray}
and
\begin{equation}
\bar{K}=\frac12\left(\frac{df}{dx}\right)^2+\frac12\left(\frac{dh}{dx}
\right)^2+\frac1{x^2}\left(\frac{du}{dx}\right)^2 \ .
\end{equation}

Moreover, using the definition (\ref{N}), Eq. (\ref{xAN}) can be
further brought to the form
\begin{equation}
\label{mu}
\mu' = \half \gamma^2 x^2 (N \bar K + (\bar U - \frac{q^2}{x^2})) \ .
\label{finiteeng}
\end{equation}

%%%%%%%%%%%%%%%%%%%%%%%%%%
\section{Exotic defects and changed asymptotic values}
%%%%%%%%%%%%%%%%%%%%%%%%%%%
As explained earlier (see also \cite{bbh}), the total 
rescaled potential
\begin{eqnarray}
{\hat V} &=& \biggl[ \frac{\lambda_1}{4}\left(\phi^a\phi^a-\eta_1^2\right)^2
+\frac{\lambda_2}{4}\left(\chi^a\chi^a-\eta_2^2\right)^2+
\frac{\lambda_3}{2}\left(\phi^a\phi^a-\eta_1^2\right)
\left(\chi^a\chi^a-\eta_2^2\right) \biggr] \frac{1}{e^2 \eta_1^4}\nonumber \\
&=&
\frac{\beta_1}{4}(h^2-1)^2+\frac{\beta_2}{4}(f^2-q^2)^2
+\frac{\beta_3}2
(h^2-1)(f^2-q^2)
\end{eqnarray}
has 
different properties according to the 
sign of $\Delta = \lambda_1 \lambda_2 - \lambda_3^2$ (which equals the sign of ${\hat \Delta}:=\beta_1 \beta_2-
\beta_3^2$).

For $\Delta > 0$ the potential is semi-positive definite and its
minima are attained for $h^2=1$ and $f^2=q^2$. In this case, requiring
that the Higgs and Goldstone field go to their respective vacuum
values ensures that the asymptotic value of the potential energy 
density is zero.

For $\Delta <0$ and $\beta_3>0$, the configuration  $h^2=1$ and $f^2=q^2$
becomes a saddle point and the two new minima occur for
\be
\label{delta1}
h^2 = 0 \ , \ \ f^2 = 
\left(q^2 + \frac{\beta_3}{\beta_2}\right)=f_{\rm min}^2 
\ee
and
\be
\label{delta2}
f^2 = 0 \ , \ \ h^2 = \left(1 
+ \frac{\beta_3}{\beta_1 } q^2 \right)= h_{\rm min}^2  .
\ee

For the case $\Delta=0$ there is a whole family of minima for
\be
h^2=1-\frac{\beta_3}{\beta_1}\left(f^2-q^2\right) \quad {\rm or}, \ {\rm 
equivalently}, \quad \sqrt{\beta_1} (h^2 -1) = \pm\sqrt{\beta_2} (q^2 - f^2)
\label{all}
\ee
where the $\pm$ corresponds to the sign of $\beta_3$.

The boundary conditions used  previously in the literature
(in \cite{bbh} for instance)

\begin{equation}
\label{bc1}
u(x=0)=1 \ , \ \ f(x=0)=h(x=0)= \mu(x=0)= 0
\end{equation}
and
\begin{equation}
\label{bc1b}
u(x=\infty)= 0 \ , \ A(x=\infty) = 1
\end{equation} 
\begin{equation}
\label{bc2}
h (x=\infty)=  1 \ ,  \ \ f(x=\infty)=  q 
\end{equation} 
are clearly not suitable for the range of possibilities we encounter in the present case.
Therefore, we adopt a different strategy: instead of fixing the values of the fields $f$ and $h$ at infinite,  
we impose that the {\bf derivative with respect to $x$}
of the fields $f(x)$ and $h(x)$ should vanish at infinity.  The system is left ``free to choose'' the asymptotic values 
of the fields. 
We then confirm that the values chosen are the correct ones.

Our boundary conditions are thus equal to the ones given above, 
except for the values of $h$ and $f$ at infinity:
the boundary conditions are   
given by conditions (\ref{bc1}), (\ref{bc1b}), but conditions (\ref{bc2}) are replaced by
\begin{equation}
\label{newbc}
h'(x)\vert_{x=\infty}=0 \ \ , \ \ f'(x)\vert_{x=\infty}=0 \ .
\end{equation}

As we confirmed numerically, the asymptotic behaviours for the fields $h$ and $f$ for $\Delta\leq 0$ are the ones 
given in (\ref{delta2}) (see below for more details).  We could also impose $u'(x)\vert_{x=\infty} = 0$ instead of $u(x=\infty) = 0$.

%%%%%%%%%%%%%%%%%%%%%%%%%%%%%%%%%%%%%%%%%%%%%%%%%%%%%%%%%%%
\subsection{From deficit angle to cosmological constant}
%%%%%%%%%%%%%%%%%%%%%%%%%%%%%%%%%%%%%%%%%%%%%%%%%%%%%%%%%%%%

The different asymptotic values of the scalar fields have drastic
consequences on the metric of the system.  For $\Delta>0$, the vacuum
values of $f$ and $h$ ($f^2=q^2$, $h^2=1$) correspond to zero
potential, whereas for $\Delta<0$ the minimum value of the potential
is negative and results in a cosmological constant. The deficit angle
of the global monopole is also tied to the
asymptotic values of the scalars.  Therefore, we find different types
of space-times in this model depending on the value of $\Delta$ and the sign of $\beta_3$. 

If $\Delta>0$ (irrespective of the sign of $\beta_3$), from (\ref{mu})
we have \be \mu'\vert_{x>>1} \sim \half \gamma^2 x^2
\left(\frac{1}{2x^4}\right) \ee leading to \be N(x>>1) \sim 1 -
\gamma^2 q^2 - 2\frac{\mu_{\infty}}{x} + \frac{\gamma^2}{2 x^2}+
O\left(x^{-3}\right) \ee with $\mu_{\infty}$ being the integration
constant.  This represents a space-time with deficit solid angle equal
to $\gamma^2 q^2$ (times $4\pi^2$) and magnetic charge equal to unity (in units of
$2\pi/e$).

However, for $\Delta<0$ and $\beta_3>0$  the fields behave asymptotically like
(\ref{delta1}) or (\ref{delta2}). We found numerically that the preferred solution is given by
(\ref{delta2}), since that corresponds to setting the global monopole (which has divergent energy) to zero.

Thus, we find the following behaviour for the metric
\begin{equation}
\mu'\vert_{x>>1} \sim \frac{\gamma^2}{2} x^2 \frac{q^4}{4}\left
(\beta_2-\frac{\beta_3^2}{\beta_1^2}\right)-
\frac{\gamma^2}{2} q^2 + \frac{\gamma^2}{4 x^2}
\end{equation} 
which gives for $N(x)$ :
\begin{equation}
\label{ads}
N(x>>1) \sim 1- 2\frac{\mu_{\infty}}{x}+ \frac{\gamma^2}{2x^2}+ 
\frac{\gamma^2}{2} q^4 
\left(\frac{\beta_3^2}{\beta_1}-\beta_2^2\right)\frac{ x^2}{6}=
1- 2\frac{\mu_{\infty}}{x} + \frac{\gamma^2}{2x^2}-  
\frac{\gamma^2 q^4 \Delta}{\beta_1^2}\frac{ x^2}{12}  \ .
\end{equation}
Note that the coefficient of the $x^2$ term is identified with
(minus one third of) the cosmological constant. This metric thus
represents an Anti-de-Sitter space-time without deficit angle and with
negative cosmological constant  $\Lambda$: \be
\Lambda:=\frac{\gamma^2 q^4\Delta}{4\beta_1^2} \ee

For $\Delta=0$ and $\beta_3>0$, the asymptotic value of $f$ and $h$ follow equation (\ref{all}), so we find that
\bea
& & \mu'\vert_{x>>1} \sim 
\half \gamma^2 (f^2-q^2)  +\frac{\gamma^2}{4x^2} \qquad \longrightarrow \nonumber\\
& & N(x>>1) \sim 1- \gamma^2 q^2 + \frac{\gamma^2}{2x^2}   
- \gamma^2 (f^2-q^2) - 2 \frac{\mu_{\infty}}{x}
\label{flat}
\eea

Out of all the possible solutions (\ref{all}), the system chooses the
one with $f=0$, since that minimises the energy, so we have (using the
definition of $\gamma$ and $q$) that the solid deficit angle cancels
out, and the metric just represents an asymptotically flat metric
without deficit angle, a Reissner-Nordstr\"om space-time.

For $\Delta=0$ and $\beta_3<0$ the solutions depend on the parameters. For $\sqrt{\beta_1} > \sqrt{\beta_2} q^2$ the situation
is analogous to the case with $\beta_3>0$ (\ref{flat}).

However, for $\sqrt{\beta_1} < \sqrt{\beta_2} q^2$ the solution is
 $h=0$ and $f^2=q^2-\sqrt{\frac{\beta_1}{\beta_2}}$, so the metric
functions are now:
\bea
& & \mu'\vert_{x>>1} \sim 
\frac{\gamma^2}{4x^2}-\half \gamma^2 \sqrt{\frac{\beta_1}{\beta_2}}
\longrightarrow \nonumber\\
& & N(x>>1) \sim 1 - \gamma^2 \left(q^2-\sqrt{\frac{\beta_1}{\beta_2}}\right)+ \frac{\gamma^2}{2x^2}   
- 2 \frac{\mu_{\infty}}{x}
\eea

This corresponds to a space-time with deficit angle 
$\propto \left(q^2-\sqrt{\frac{\beta_1}{\beta_2}}\right)$
and magnetic charge equal to unity.

These cases will be illustrated by numerical simulations in section \ref{numerical}.

%%%%%%%%%%%%%%%%%%%%%%%%%%%%%%%%%%%%%%%%%%%%%%%%%%
\subsection{From higgsed phase to symmetric phase}
%%%%%%%%%%%%%%%%%%%%%%%%%%%%%%%%%%%%%%%%%%%%%%%%%%

The AdS solution obtained in the previous section for $\Delta<0$ and $\beta_3>0$ was a consequence of the new boundary conditions given by 
equations (\ref{bc1}), (\ref{bc1b}) and (\ref{newbc}). But one could also wonder if that is the best choice of boundary conditions.

We have to bear in mind that we are actually analysing {\it static} solutions for the problem given some values of the parameters. But we could 
try to imagine a situation in which we begin with some fixed values of the parameters, and $\beta_3$ varies from $\beta_3=0$ to higher values; which would correspond to 
$\Delta$ ranging from $\Delta>0$ to $\Delta<0$ passing through $\Delta=0$. 

However, we have seen that at $\Delta=0$, the global monopole disappears. Therefore, the Ansatz for the global scalar function together with its boundary conditions
might not be optimal for this case. Let us consider the case where the conditions for $\chi$ are changed to an Ansatz with no winding
\be
\chi^a=\eta_1 f(r)
\ee
The equation of motion for $f$ (\ref{eqnf}) will change to
\begin{equation}
\left[x^2AN f'\right]'=
A\left[x^2\beta_2 f(f^2-q^2)+
{x^2\beta_3 } (h^2-1)f\right] 
\end{equation}

There is no need of imposing $f(0)=0$ now, so we can change that to $f'(0)=0$ also. We then  obtain a different solution for the system when $\Delta<0$. It can be seen analytically that the configuration
\be
f(x)=f_{\rm min}\,,\quad
h(x)=0\,,\quad
u(x)=1\,,\quad
\mu(x)=0
\ee
is  a solution.
We could have used this boundary condition from the beginning, and all the solutions obtained so far hold.) 
This corresponds to having all the windings disappear, the Higgs mechanism ceases to operate, and we are left in the true vacuum.
The value of the constant $f_{\rm min}$ is the one that minimizes the potential (\ref{delta1}). 

Once the Higgs field has vanished everywhere, the Ansatz and boundary conditions for $\phi$ should also be revised, and changed as in the $\chi$ case. So the system would in principle choose the solution corresponding to the lowest minimum (\ref{minimum1}) or (\ref{minimum2}).

Of course, the question arises of how these processes would take place in a dynamical setting, and whether the system would evolve toward one of these solutions getting rid of all the topology altogether. In physical terms one can imagine the formation of a condensate of the global scalar field whose effect is to lower the value of $f$ and expand the magnetic core. Once $f$ is zero in a sufficiently large region a condensate of $\phi$ with no winding can develop at the expense of the global condensate and eventually reach the solution with $f =f_{\rm min}$, $h=0$.
We will investigate this process in a subsequent publication.

%%%%%%%%%%%%%%%%%%%%%%%%%%%%%%%%%%%%%%%%%%%%%%%
\section{Numerical results}
\label{numerical}
%%%%%%%%%%%%%%%%%%%%%%%%%%%%%%%%%%%%%%%%%%%%%%%

We have solved the equations (\ref{eqnf})-(\ref{eqnA}) subject to the
boundary conditions (\ref{bc1}), (\ref{bc1b}) and (\ref{newbc}) for
different values of the parameters using relaxation methods and the
damped Newton method of quasi-linearization \cite{acr}.

\begin{figure}[!htb]
\centering
\leavevmode\epsfxsize=9.0cm
\epsfbox{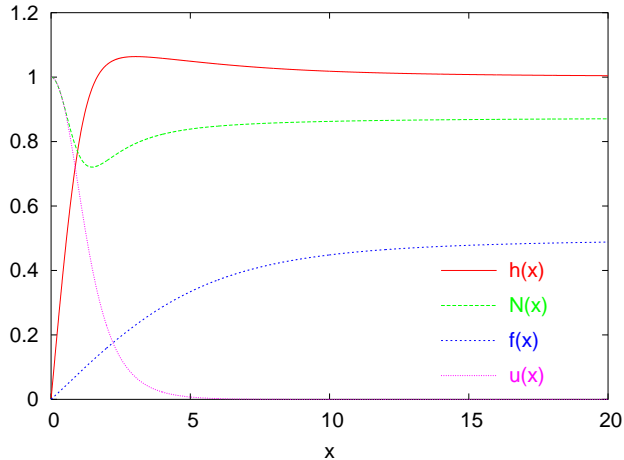}\\
\caption{\label{figold} The profile of the fields $f(x)$, $h(x)$, $u(x)$ and $N(x)$ for the composite
monopole with $\beta_1=\beta_2=1$, $\beta_3=0.75$ (thus, $\Delta\sim 0.44$), 
$q=0.5$ and $\gamma^2=0.5$. 
}
\end{figure}

We first reproduced known results \cite{spi,bbh} for the case
$\Delta>0$. Despite using different boundary conditions, the system
finds the same solution as in previous work \cite{spi,bbh}.  Figure
\ref{figold} shows the typical type of profile obtained for the case
$\Delta>0$: the Higgs field $h$ grows higher than its asymptotic value
and then goes down to $1$. The global field $f$ tends to its vacuum
value $q$ more slowly than in a non-coupled case.

\begin{figure}[!htb]
\centering
\leavevmode\epsfxsize=9.0cm
\epsfbox{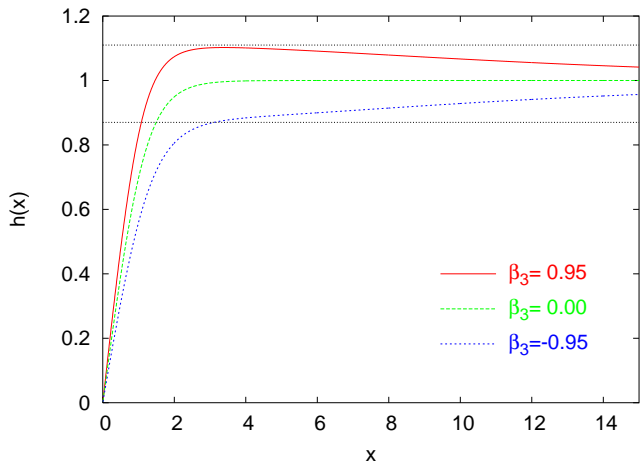}\\
\caption{\label{lambdaneg} 
Profiles of the function $h$ for $\beta_1=\beta_2=1$, $q=0.5$, $\gamma^2=0.5$ and different values of $\beta_3$. The horizontal lines shown correspond to
$h_{\rm min}(\beta_3=0.95)\sim1.11$ and $h_{\rm min}(\beta_3=-0.95)\sim0.87$ given in equation (\ref{delta2}). Note that close to the core, the function
tends to that minimum. 
}
\end{figure}

In Figure \ref{lambdaneg} we show the profiles of the function $h$ for different values of $\beta_3$, always in the
regime $\Delta>0$. In all those cases, the profile of $f$ does not vary much (not shown). Comparing to the non-interacting
case $\beta_3=0$, 
the function $h$ goes to higher (lower) values around the core for $\beta_3$ positive (negative). The reason could be
that the system ``sees'' the new minimum $(f=0,h=h_{\rm min})$, and in places where $f<<q$, it tries to go to that minimum. Moreover, around the 
core of the defect the metric function is usually small, and thus, gradient energy becomes ``cheaper'' than potential energy.

On the other hand, when $\Delta \leq 0$ $(\beta_3>0)$ the solutions are very different: the 
function $f(x)\equiv 0$ and the function $h(x)$ tends asymptotically
to the value given in (\ref{delta2}), i.e. the global monopole disappears and 
the Higgs field asymptotes 
to its new vacuum expectation value. This happens irrespective of the values of the parameters $q$, $\beta_1$ and $\beta_2$, even if $V_h<V_f$.
This is the case also for $\Delta=0$, 
where all the possible minima are degenerate in {\it potential} energy; but it is obviously favourable to set the global monopole to zero, 
since it has divergent gradient energy.

\begin{figure}[!htb]
\centering\leavevmode\epsfxsize=7.5cm \epsfbox{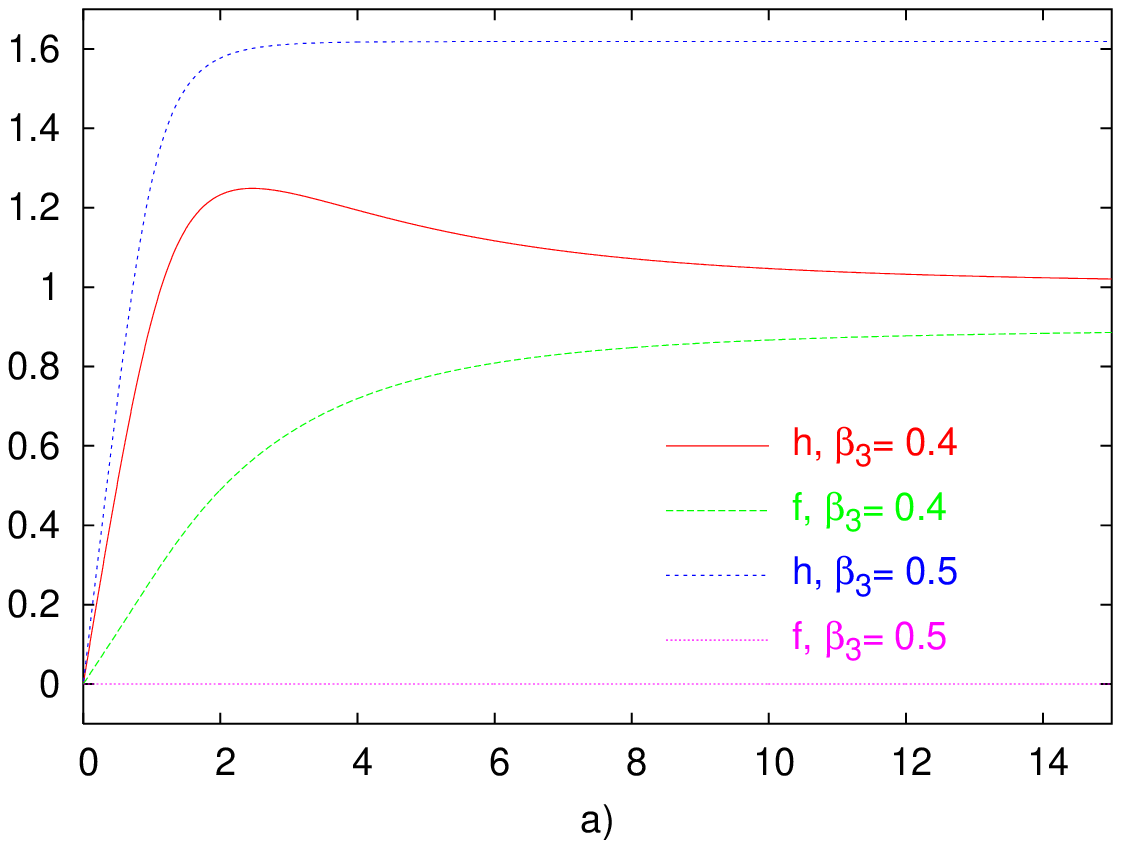} 
\hspace{.9cm}\leavevmode\epsfxsize=7.5cm \epsfbox{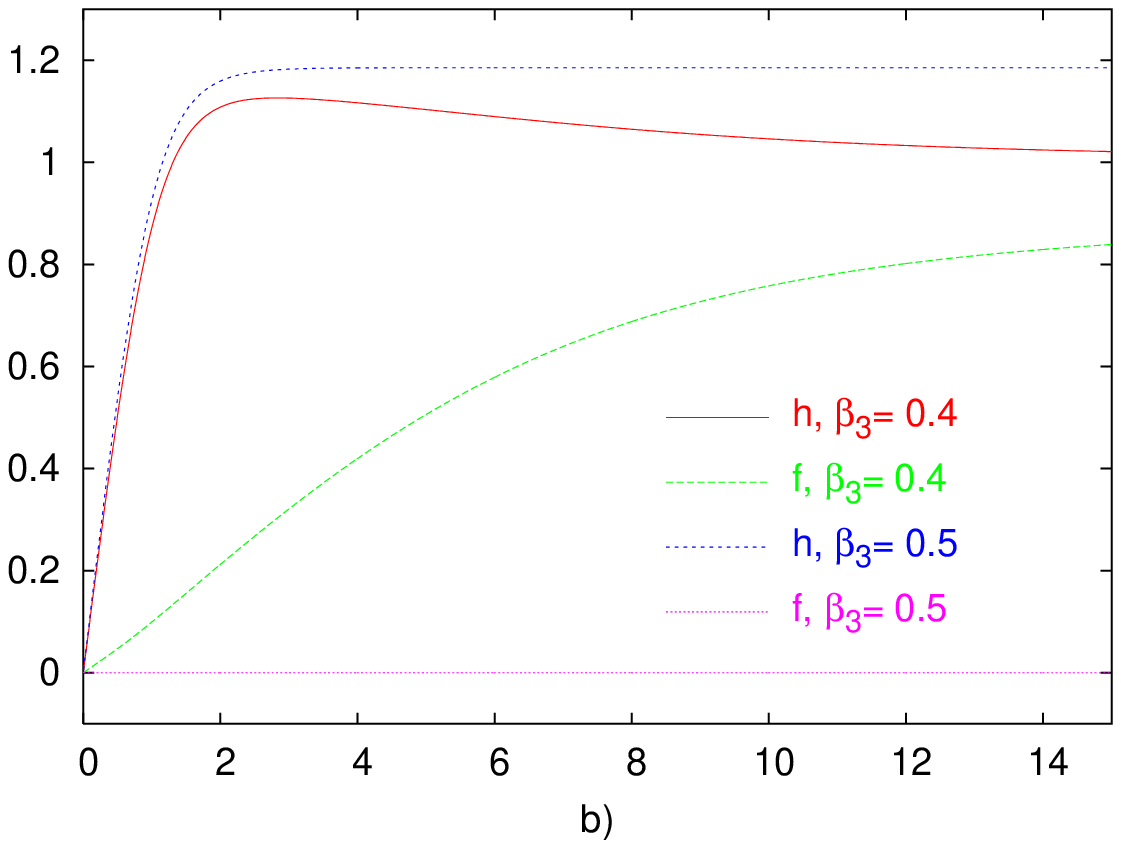}\\
\vspace*{12pt}
\caption{{\label{global0}}
a) Profile of the functions $f$ and $h$ for different values of $\beta_3$, with $\beta_1=0.25, \beta_2=1$, $q=0.9$ and $\gamma^2=0.5$. In this case
$\frac{\beta_1}{\beta_2}<q^4$ \qquad b) Same as before but with $\beta_1=1$ and $\beta_2=0.25$, so that $\frac{\beta_1}{\beta_2}>q^4$.
Note that in both cases, the global field $f$ is the one that vanishes when $\Delta=0$.}
\end{figure}

\begin{figure}[!htb]
\centering\leavevmode\epsfxsize=7.5cm \epsfbox{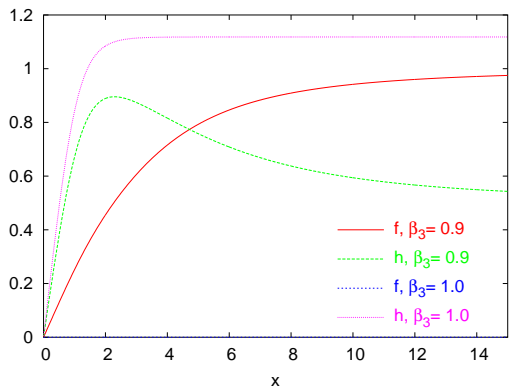} 
\vspace*{12pt}
\caption{{\label{global1}} Profile of functions $f$ and $h$ for $q>1$, showing
that the global monopole vanishes
($\beta_1=1$, $\beta_2=1$, $\gamma^2=0.5$, $\eta_1=0.5$, $\eta_2=1$).}
\end{figure}

In order to show that the global monopole is the one that disappears,
we include Figures \ref{global0} and \ref{global1}. In Figure
\ref{global0} a) and b) we plot the functions $f$ and $h$ for
different values of the parameters. One set of curves in each figure
corresponds to $\Delta>0$ (to have the composite defect as a
reference) and the other set to $\Delta=0$. The case with $\Delta<0$
is analogous to $\Delta=0$. Figure \ref{global0} a) has
$\sqrt{\frac{\beta_1}{\beta_2}}<q^2$ and b) has
$\sqrt{\frac{\beta_1}{\beta_2}}>q^2$. In Figure \ref{global1} we
present a case with   $q>1$ ($\eta_1=0.5$ and $\eta_2=1$) to
show that the global monopole also disappears in this case.

\begin{figure}[!htb]
\centering
\centering\leavevmode\epsfxsize=7.5cm \epsfbox{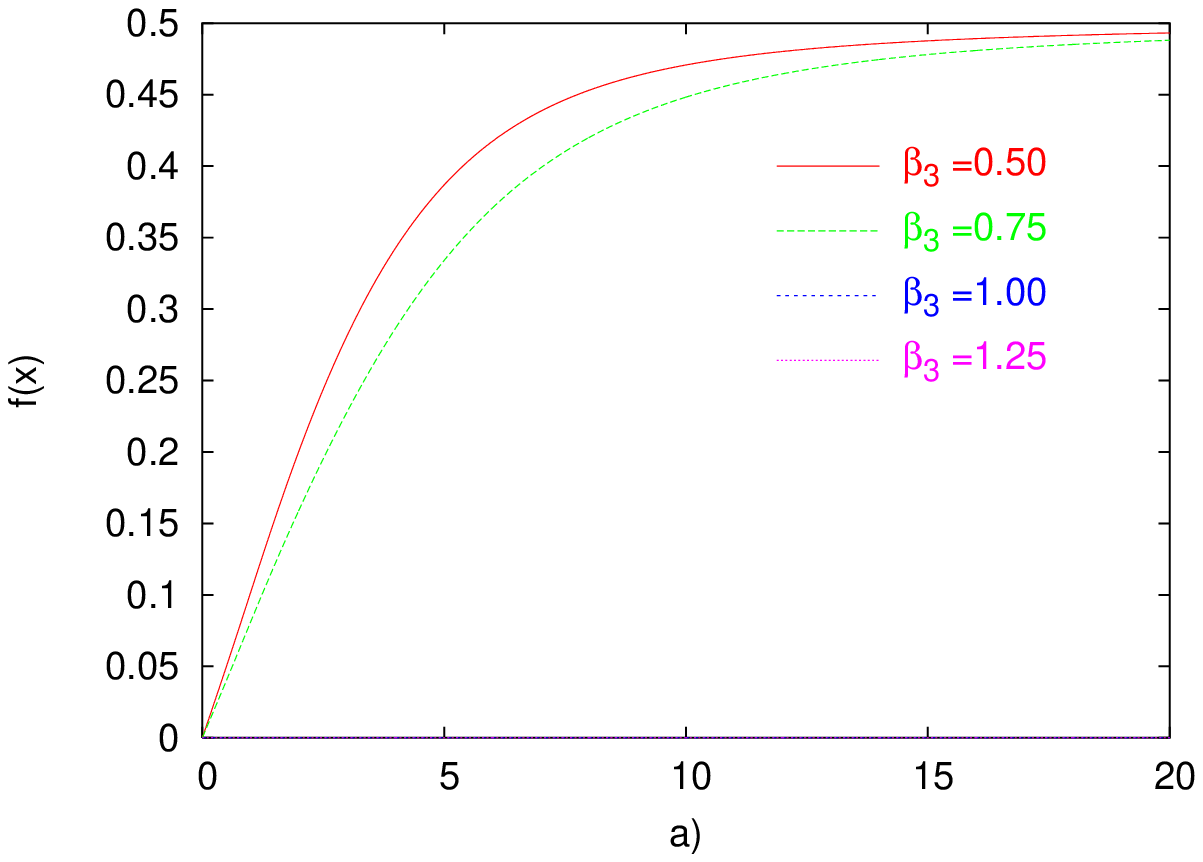} 
\hspace{.9cm}\leavevmode\epsfxsize=7.5cm \epsfbox{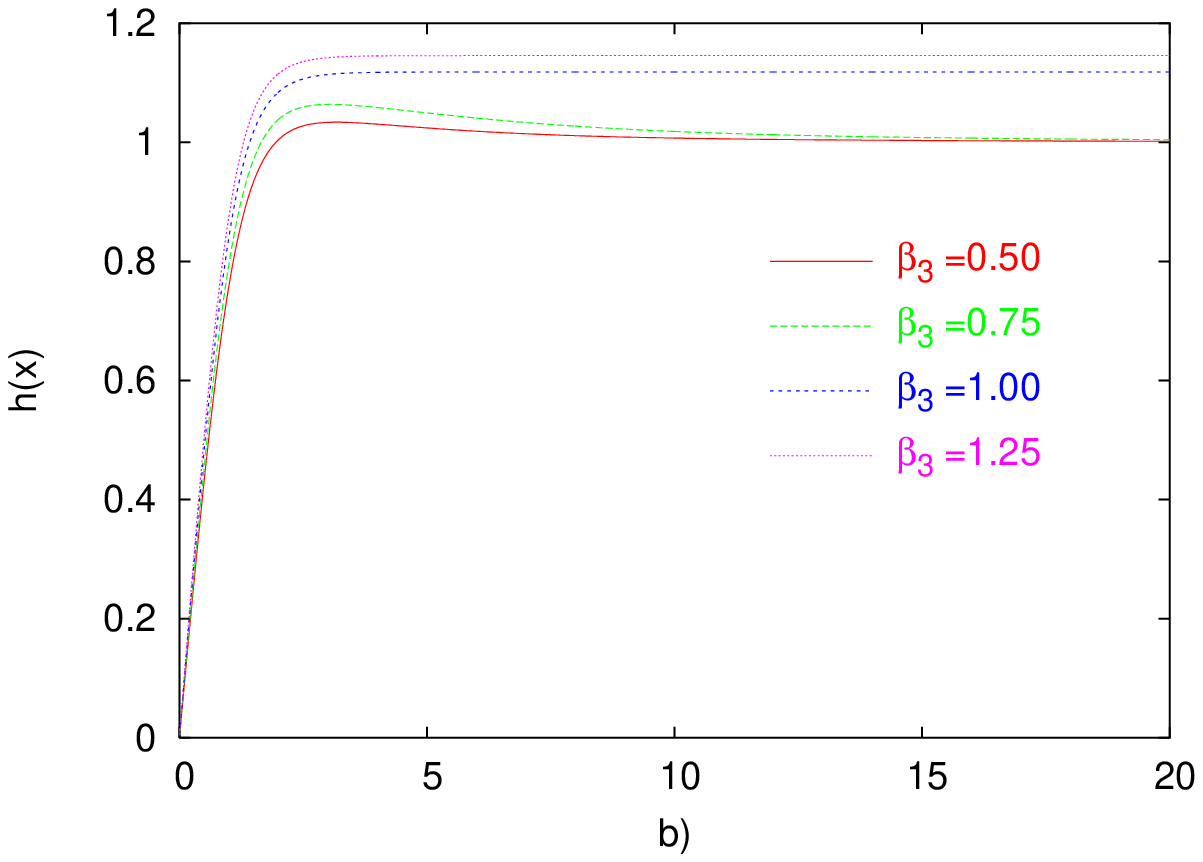}\\
\caption{\label{fig1} The profile of the global field function $f(x)$ (a) and the Higgs field 
$h(x)$ (b)
of the composite monopole system is shown
for $\beta_1=\beta_2=1$, $q=0.5$, $\gamma^2=0.5$ and different values of $\beta_3$. Note that 
$\Delta > 0$ for $\beta_3=0.5$ and $\beta_3=0.75$, $\Delta=0$ for $\beta_3=1$ and
$\Delta < 0$ for $\beta_3=1.25$. The profile $f(x)\equiv 0$ for $\Delta\leq 0$. The profile of 
$h(x)$ also changes with $\Delta$: for $\Delta>0$ it raises beyond its asymptotic value and 
goes down; for $\Delta \leq 0$, it goes to its asymptotic value monotonically.}
\end{figure}

\begin{figure}[!htb]
\centering
\leavevmode\epsfxsize=9.0 cm
\epsfbox{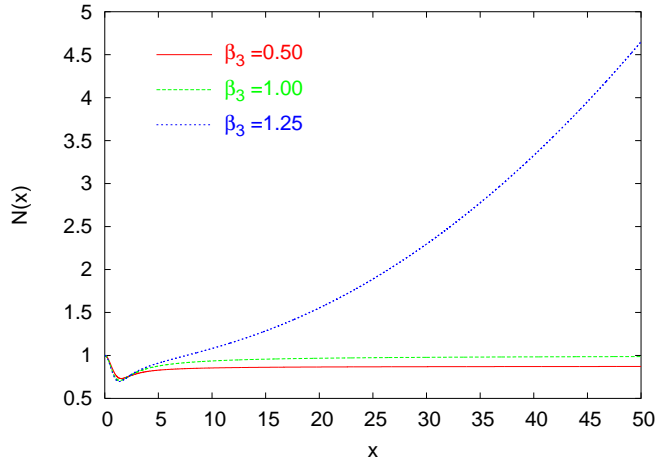}\\
\caption{\label{fig3} Same as Figure \ref{fig1} for the metric function
$N(x)$ of the composite monopole system. The powerlaw
increase for $\beta_3>1$ signals a cosmological constant}
\end{figure}

Furthermore, the metric function $N(x)$
asymptotically tends to the Anti-de-Sitter form given in (\ref{ads}).
This is demonstrated in Figure \ref{fig1} and Figure \ref{fig3} for $\beta^2_1=\beta^2_2=1$,
$\gamma^2=0.5$, $q=0.5$ and different values of $\beta_3$. Clearly for $\Delta < 0$ ($\beta_3>1$ in this case), $f(x)\equiv 0$
and $h(x)$ tends asymptotically to $h_{\rm min}$ which here is $h_{\rm min}\approx 1.12$ for 
$\beta_3=1$ and $h_{\rm min}\approx 1.15$ for $\beta_3= 1.25$. Moreover, as can be
seen from Figure \ref{fig3}, the metric function tends to a value smaller than
one for $\Delta > 0$, thus representing a space-time with deficit angle.
For $\Delta=0$, the space-time becomes asymptotically flat, while for $\Delta < 0$, the
metric function $N(x)$  rises rises following a power law representing asymptotically an 
Anti-de-Sitter--Reissner-Nordstr\"om (AdSRN) space-time.

\begin{figure}[!htb]
\centering
\leavevmode\epsfxsize=9.0 cm
\epsfbox{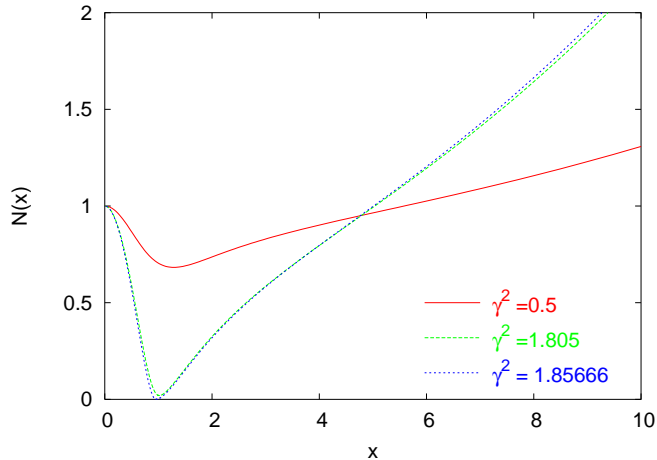}\\
\caption{\label{fig4} 
The metric function $N(x)$ of the composite monopole
system is shown for $\beta_1=\beta_2=1$, $\beta_3=1.25$ (thus $\Delta < 0$), $q=0.5$ and increasing values of $\gamma$. }
\end{figure}

When increasing $\gamma$, the minimum of the metric function gets deeper and finally
at some $\gamma=\gamma_{cr}(q,\beta_i)$ a degenerate horizon forms at $x=x_h$ with $N(x_h)=0$,
$N'(x)\vert_{x=x_h}=0$.
This is demonstrated in Figure \ref{fig4} for $\beta_1=\beta_2=1$, $\beta_3=
1.25$, $q=0.5$
and increasing $\gamma$. At the same time the matter functions become equal to
their vacuum values outside of this horizon, 
i.e. $h(x > x_h)\equiv h_{\rm min}$ and $A(x > x_h)\equiv 1$ with an infinite derivative
at $x=x_h$.

The solution thus represents an extremal 
Anti-de-Sitter--Reissner-Nordstr\"om (AdSRN) 
black hole for $x > x_h$  with
\begin{equation}
N(x > x_h)= 1- 2 \frac{\mu_{\infty}}{x}- \Lambda \frac{x^2}{3}+\frac{\gamma^2}{2x^2} 
\end{equation}
and extremal horizon at
\begin{equation}
x_h=\frac{1}{\sqrt{2\vert\Lambda\vert}}\sqrt{-1+
\sqrt{1-2\gamma^2\Lambda}} \ \ \ , \ \ \ 
\end{equation}
while it is non-trivial and non-singular for $0 \leq x < x_h$.
For $\beta_1=\beta_2=1$, $\beta_3=1.25$, $q=0.5$, $\gamma^2= \gamma_{cr}^2\approx 1.85666$, we find
from the above formula that $x_h\approx 0.94$ which agrees very well with our
numerical results.

\begin{figure}[!htb]
\centering\leavevmode\epsfxsize=7.5cm \epsfbox{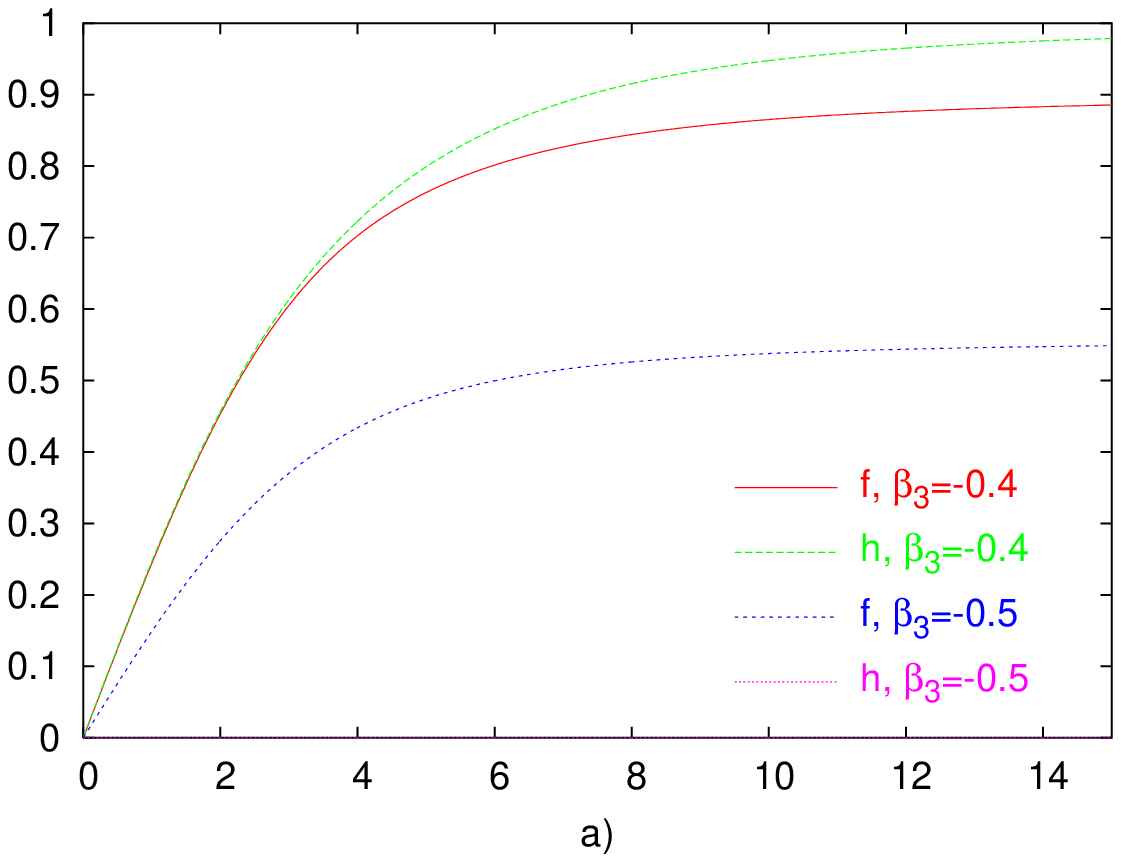} 
\hspace{.9cm}\leavevmode\epsfxsize=7.5cm \epsfbox{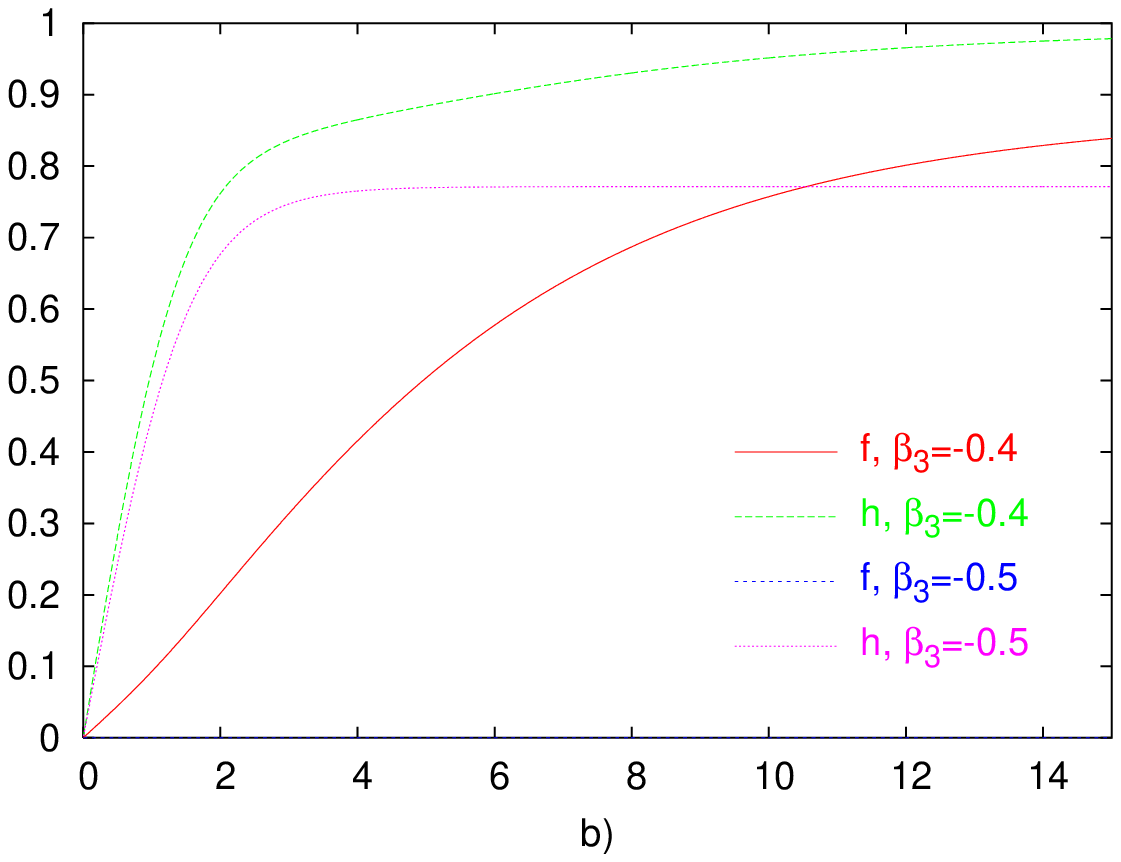}\\
\vspace*{12pt}
\caption{{\label{globalneg}}
This is the equivalent of Figure \ref{global0} but with $\beta_3<0$. Contrary to that case, depending on the value of the parameters
the local monopole $f$ (Figure a, $\frac{\beta_1}{\beta_2}<q^4$) or the global one $h$ (Figure b, $\frac{\beta_1}{\beta_2}>q^4$) will disappear.
}
\end{figure}

The case with $\Delta=0$ and $\beta_3<0$ is special. As mentioned above, one or the other (or both) 
monopoles will disappear depending on the values of the parameters. Thus, for the case $\frac{\beta_1}{\beta_2}>q^4$ the global monopole will disappear (see Figure (\ref{globalneg}) b), as in the case
with $\beta_3>0$. But for $\frac{\beta_1}{\beta_2}<q^4$, the local one disappears (see Figure (\ref{globalneg}) a), even if the global one has divergent energy. This is due to the fact that
the configuration with $f=0$ ceases to be a solution, so the system does not have the choice of making the global monopole disappear. There are no solutions for $\beta_3<0$ and $\Delta<0$ since the potential is  not bounded from below.

%%%%%%%%%%%%%%%%%%%%%%%%%%%%%%%%%%%%%%%%
\section{Summary and discussion}
%%%%%%%%%%%%%%%%%%%%%%%%%%%%%%%%%%%%%%%%

Defects containing global and local symmetries have recently become an
interesting arena in which one can try to model some of the more
exotic properties of superstrings and branes within a field theory
context. This is particularly important for applications to cosmology.

In this paper  we revisited  a simple case, a composite global/local
monopole with a quartic interaction term between the two scalar
triplets.  The static solutions in this model were analysed in
\cite{spi,bbh} for weak coupling, here we have investigated the strong
coupling case and discovered a dramatic change in the nature of the
solutions.

There is a critical value of the coupling for which both the magnetic
monopole and the global monopole become non-topological and
potentially unstable. If the interaction coupling constant is positive
(corresponding to a repulsive interaction between the cores) we find
that the global monopole always disappears.  More precisely, for
values higher than the critical coupling, we find that the static
solution is a local monopole in Anti-de-Sitter space-time which in the
limit of sufficiently large gravitational coupling forms a degenerate
horizon.  Outside of this horizon, the solution then corresponds to an
Anti-de-Sitter--Reissner-Nordstr\"om solution, while inside it is
non-trivial and non-singular. If the interaction term has the opposite
sign (core attraction) we find that either or both monopoles can
disappear depending on parameters. 

We have discussed what are the appropriate boundary conditions in each
regime, and how to implement them in numerical calculations. We should
stress that all our solutions are in principle constrained, since the
condition of spherical symmetry forces both monopole cores to sit at
the origin, an assumption that is not always justified dynamically (but that makes  sense in the context of applications to braneworlds). It
leaves open the question of how a network of such defects would form
and evolve in a cosmological context, in particular for what ranges of
parameters the two kinds of monopoles would evolve more or less
independently and for what ranges they would be tightly coupled (which
would enhance magnetic monopole annihilation).

We have concentrated on the nature of the static solutions for various
regimes of the couplings, in particular those regimes in which the
monopoles become unstable. A dynamical study of how these defects
decay is of course very interesting in its own right but there are at
least three more reasons why the extension of these results to the
time-dependent case should be considered.

First of all, time-dependent solutions with deficit angles larger than
$4\pi^2$ can give rise to topological inflation \cite{Linde,Vilenkin}
which in this composite model would end naturally when the defects
decay (see \cite{abr} for a related proposal).

Secondly, an important change in paradigm that has resulted from
superstring cosmology and braneworlds is that coupling constants are
not truly constant, they are the expectation values of fields and may
be space-time dependent. In our particular example the time evolution
of one coupling constant (the interaction coupling) could completely
change the structure of spacetime. For low coupling the system looks
like a magnetic monopole and a global monopole superimposed, and
spacetime is asymptotically Reissner-Nordstrom with a deficit angle
and no cosmological constant. As the coupling reaches the critical
value, the global monopole disappears and with it the deficit angle,
and we are left with a magnetic monopole. The magnetic core is
unstable and develops a scalar condensate of the global field which
makes the core expand indefinitely. In this last step, the $SU(2)_{\rm
gauge}$ symmetry is restored in the expanding core and the gauge field
eventually becomes massless everywhere. The end result is a transition
from the Higgs phase to the symmetric phase.

We expect many of the qualitative features found here (decay
of deficit angles, cosmological constant reduction, symmetry
restoration) to carry over to the composite $U(1)_{\rm global} \times
U(1)_{\rm gauge}$ vortex case, where they should have an application
in braneworlds and in the (toy) modelling of orbifold and conifold
transitions.

\section{Acknowledgments}
A.A. thanks Leo van Nierop and Roland de Putter for discussions. B.H. and J.U. thank the University
of Leiden, where part of this work was done, for hospitality.
This work was partially supported by the Netherlands Organization for
 Scientific Research (NWO) under the VICI programme, the ESF Programme
COSLAB - Laboratory Cosmology, FPA 2002-02037 and 9/UPV00172.310-14497/2002. 
J.U is supported by the Spanish {\it Secretar{\'{\i}}a de Estado de Educaci\'on y Universidades} and {\it Fondo Social Europeo}.


\begin{thebibliography}{99}
\bibitem{foot} For $\lambda_3>0$ the cores repel, but even for $\lambda_3\leq0$ there may be a gravitational instability \cite{AU,B}
\end{thebibliography}
\end{document}